# Pattern Recognition of Illicit E-Waste Misclassification in Global Trade Data

MUHAMMAD SUKRI BIN RAMLI
Asia School of Business
Kuala Lumpur, Malaysia
Email: m.binramli@sloan.mit.edu

**Abstract**

The global trade in electronic and electrical goods is complicated by the challenge of identifying e-waste, which is often misclassified to evade regulations. Traditional analysis methods struggle to discern the underlying patterns of this illicit trade within vast datasets. This research proposes and validates a robust, data-driven framework to segment products and identify goods exhibiting an anomalous "waste signature" a trade pattern defined by a clear 'inverse price-volume'. The core of the framework is an Outlier-Aware Segmentation method, an iterative K-Means approach that first isolates extreme outliers to prevent data skewing and then re-clusters the remaining products to reveal subtle market segments. To quantify risk, a "Waste Score" is developed using a Logistic Regression model that identifies products whose trade signatures are statistically similar to scrap. The findings reveal a consistent four-tier market hierarchy in both Malaysian and global datasets. A key pattern emerged from a comparative analysis: Malaysia's market structure is defined by high-volume bulk commodities, whereas the global market is shaped by high-value capital goods, indicating a unique national specialization. The framework successfully flags finished goods, such as electric generators (HS 8502), that are traded like scrap, providing a targeted list for regulatory scrutiny.

1.  **Introduction**

The escalating global trade in electronic and electrical goods presents a significant challenge for environmental governance, particularly concerning the tracking of e-waste across vast geographies (Lepawsky, 2015). A primary obstacle is the intentional misclassification of hazardous e-waste under legitimate Harmonized System (HS) codes, a practice that makes enforcement of international regulations like the Basel Convention notoriously difficult (Bisschop, 2012; UN Environment Programme, 1989). This issue is not random but is a specific instance of the broader analytical challenge of anomaly detection in large, noisy datasets (Chandola, Banerjee, & Kumar, 2009), often driven by economic incentives to avoid tariffs (Bolton & Hand, 2002; Fisman & Wei, 2004). Standard analytical methods, however, often fail to systematically identify the underlying market structure or flag these suspicious trade flows. Therefore, this research aims to design and validate a pattern recognition framework capable of overcoming these challenges. Our central thesis is that an iterative clustering methodology, combined with a model-driven "Waste Score," can effectively uncover hidden market structures and quantify a distinct "waste signature" a pattern of low price, high volume, and price decay. The framework is designed to segment products into a clear market hierarchy, quantify the degree to which a product's trade pattern resembles scrap, and identify specific finished goods at high risk of being used for misclassified e-waste shipments. The anomalies identified by this framework are more than mere labelling errors. The detected signature of falling prices and high volume is a red flag for a dual-motive crime. Primarily, it facilitates sanction evasion by disguising prohibited waste as legitimate goods. Concurrently, this pattern mimics the value manipulation techniques of Trade-Based Money Laundering (TBML), a significant financial crime (Financial Action Task Force, 2021). This misuse of legitimate HS codes represents a form of deliberate data poisoning, compromising the integrity of global trade data and leading to flawed economic analysis and ineffective environmental policy.

**Figure 1. System Dynamics of Illicit Trade and Regulatory Enforcement.**

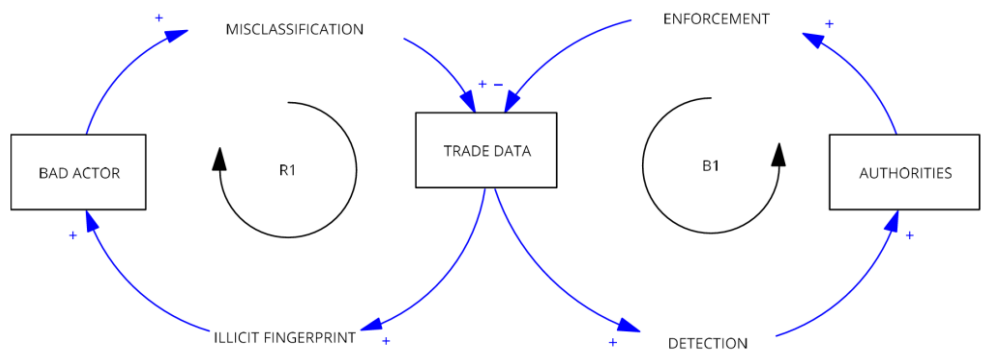

Source: Processed by Author (2025)

The relationship between illicit traders and enforcement agencies can be understood through two competing feedback loops, as illustrated in Figure 1. A reinforcing loop (R1) is driven by bad actors who misclassify goods. Paradoxically, the more they engage in this activity, the more they create a distinct 'illicit fingerprint of waste signature' within the trade data. This growing, recognizable pattern then activates a balancing loop (B1). Here, the clearer signature enables authorities to improve detection and apply targeted enforcement, which in turn counteracts the misclassification and curbs the illicit activity. The central challenge for governance is to ensure the balancing loop of enforcement becomes more effective and responsive than the reinforcing loop of crime. To analyze this dynamic in a real-world context, this study focuses on Malaysia. Malaysia is an ideal case study due to its prominent role as a global exporter of commodity metals. The country's high volume of trade on commodity metals is directed toward major industrial economies such as Japan, the USA, and China (Figure 2), and national export data reveals a dramatic and volatile increase in both trade value and volume over the past decade (Figure 4). Crucially, this trade portfolio includes many of the high-value materials commonly recovered from e-waste, such as gold, platinum, cobalt, and tantalum (Figure 3), underscoring Malaysia's significance in the global supply chain for both primary and secondary raw materials. This large-scale, legitimate trade in scrap and processed metals creates a complex environment that can be exploited to disguise materials sourced from illicit e-waste processing. This hidden activity carries dire environmental and public health consequences. The rudimentary and unsafe recycling methods often used in informal settings release a host of toxic materials such as lead, mercury, and cadmium into soil and water sources. Furthermore, the open burning of plastic components to recover metals releases carcinogenic dioxins into the atmosphere, posing severe health risks from neurological damage to cancer for both the unprotected workforce and the wider community.

**Figure 2. Malaysia's Top 10 Export Partner Countries/Areas by Number of Trade Records on Commodities Metal.**

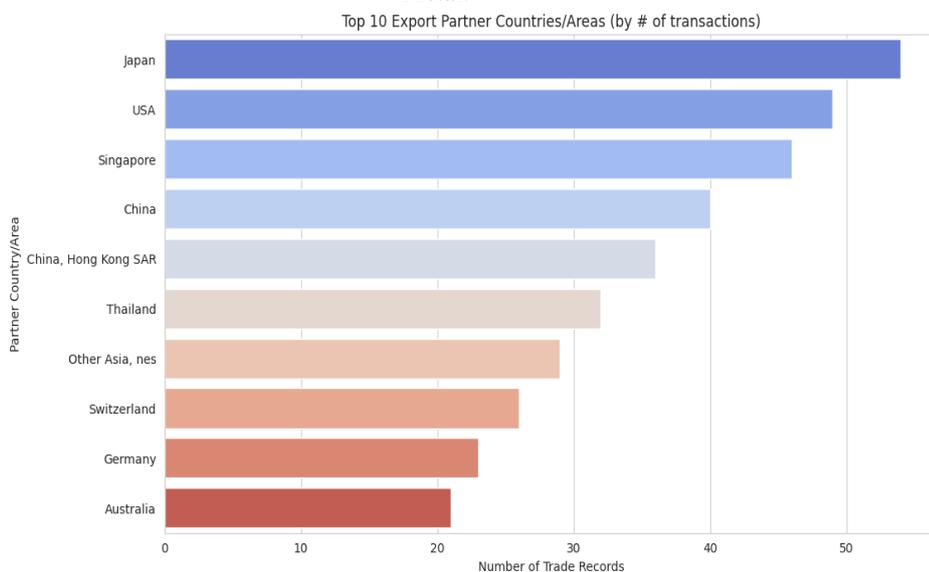

Source: Processed by Author (2025)

**Figure 3. Malaysia's Top 10 Most Valuable Export Commodities Metal by Price per kg.**

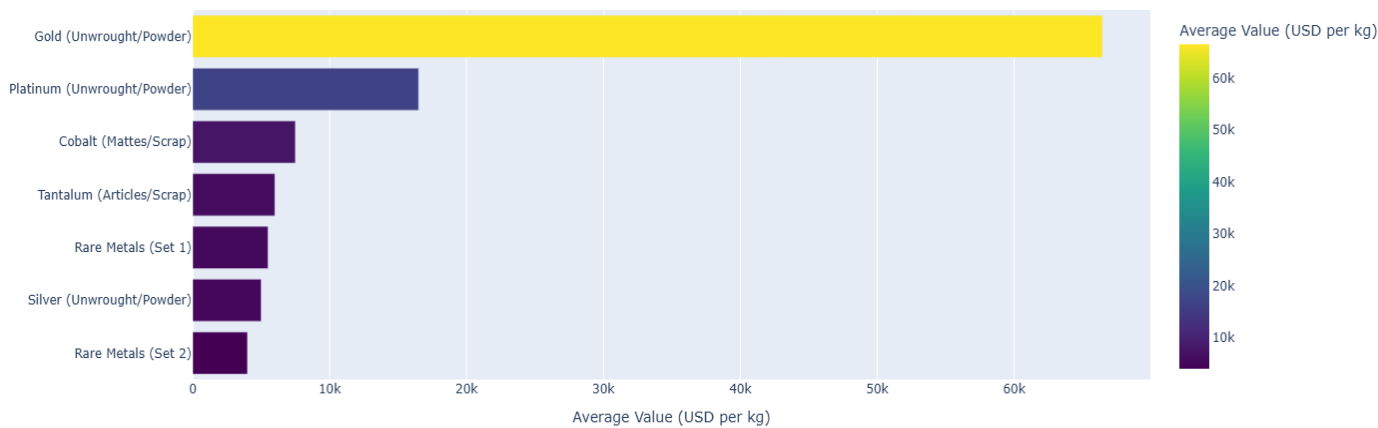

Source: Processed by Author (2025)

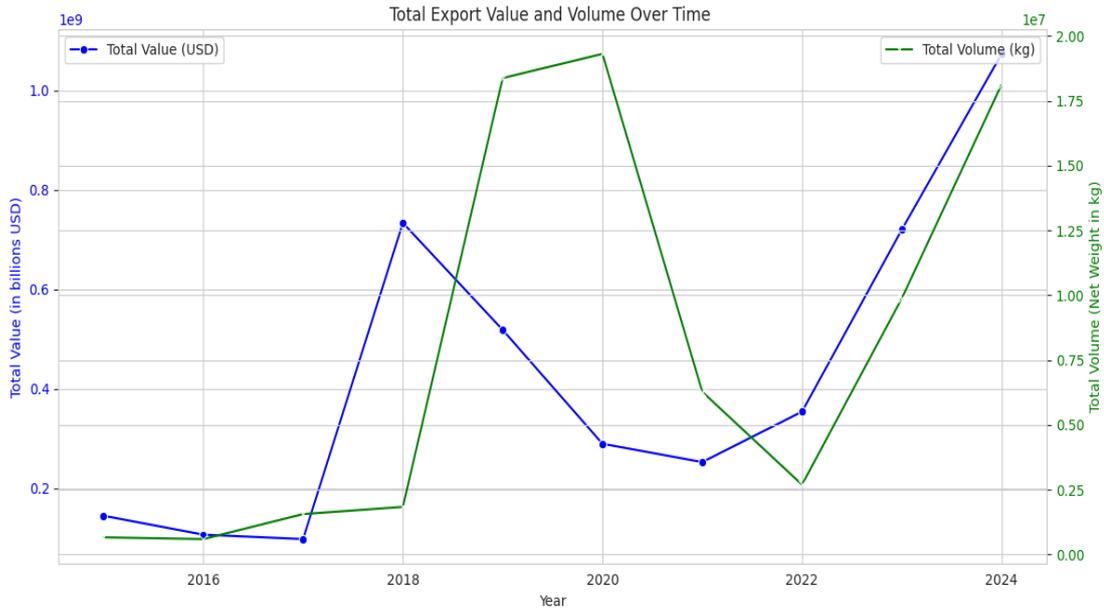

Figure 4. Malaysia's Total Export Value (USD) and Volume (kg) Over Time on Commodities Metal.

Source: Processed by Author (2025)

To identify products with trade signatures indicative of waste, we developed a structured, four-stage analytical pipeline. This end-to-end framework is designed to move from raw data to actionable intelligence in a transparent, reproducible, and statistically robust manner (Popper, 2002). The entire process, visualized in the methodological flowchart, ensures that each step logically builds upon the last, culminating in a practical tool for regulatory oversight. The workflow was executed in Python, primarily using the scikit-learn library (Pedregosa et al., 2011).

## 2. Methodology

### 2.1 Stage 1: Data Preparation and Feature Engineering

The process began with data preparation and feature engineering, a critical step for transforming raw trade data into informative variables for machine learning (Zheng & Casari, 2018). Data Sourcing and Cleaning The analysis is based on public UN Comtrade records for six-digit Harmonized System (HS) codes within Chapter 39, covering the period 2020–2024. This raw data underwent rigorous preparation to ensure quality and comparability. First, we performed unit harmonization, standardizing all trade volumes to kilograms and monetary values to USD. Missing values were handled by interpolating short gaps and excluding codes with more than 20% of their annual records missing. All monetary figures were adjusted for annual inflation, and extreme outliers were mitigated by capping values at the first and ninety-ninth percentiles. Feature Engineering From this cleaned and harmonized dataset, we engineered the foundational metric of unit price (USD per KG). Based on this, we generated five primary features to capture the economic behavior of each product code. To model more complex dynamics, we supplemented these with interaction terms and log-transformations to reduce skewness. Finally, all engineered features were standardized via z-score normalization before modeling to ensure equal weighting and facilitate algorithmic convergence. The core engineered features are detailed in Table 1.

Table 1. Engineered Features for Market Analysis

| Feature Name | Description | Rationale / Purpose |
|---|---|---|
| avg_kg | Average annual traded volume in kilograms. | Measures market size and scale. |
| avg_price | Average annual unit price (USD per KG). | Captures the economic value of the commodity. |
| price_volatility | Standard deviation of the annual unit price. | Quantifies market instability and price risk. |
| kg_trend | Slope of linear regression on trade volume. | Indicates growth (>0) or decline (<0) in volume. |
| price_trend | Slope of linear regression on unit price. | Indicates price inflation (>0) or deflation (<0). |

Source: Processed by Author (2025)

**2.2 Stage 2: Outlier-Aware Segmentation**

At the core of our study is a novel iterative segmentation approach. An initial K-Means clustering run, with the number of clusters informed by the Elbow Method, revealed that a few extreme outliers disproportionately skewed the results. This "outlier effect" isolated hyper-volume and hyper-value products while grouping the vast majority of goods into a single, undifferentiated cluster, justifying a more nuanced methodology. Our iterative framework first identifies and isolates these primary outliers, then re-clusters the remaining "core" products to reveal more subtle market segments. To validate this, we compared the results to the density-based algorithm DBSCAN, known for its ability to identify noise (Ester et al., 1996). Products flagged as outliers by both methods were considered "dual confirmation" outliers, strengthening confidence in their unique status. Figure 5 shows line graph plots the Within-Cluster Sum of Squares (WCSS) against the number of clusters (K). This plot provides the statistical justification for selecting the number of clusters in the K-Means algorithm. The "elbow" point, where the rate of decrease in WCSS slows, indicates the optimal trade-off between model complexity and explanatory power, preventing overfitting.

**Figure 5: Elbow Method for Determining Optimal Clusters (K)**

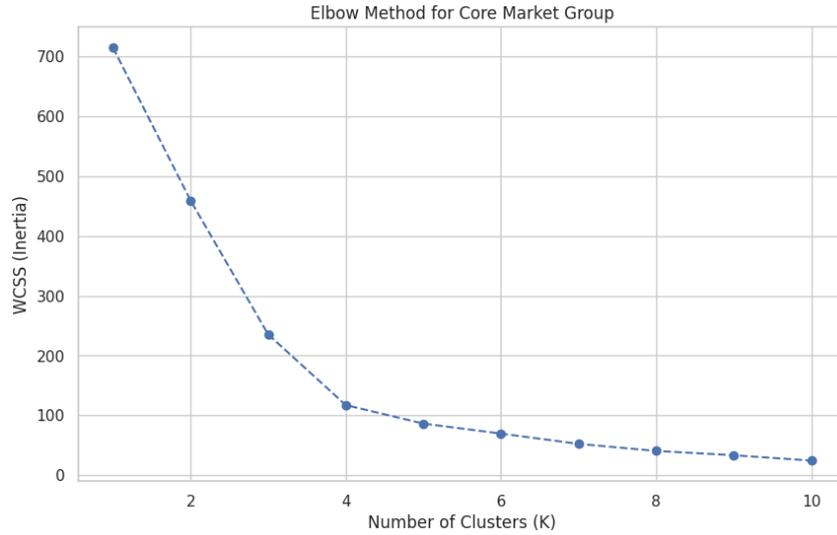

Source: Processed by Author (2025)

**Figure 6: Initial Segmentation and the Outlier Effect**

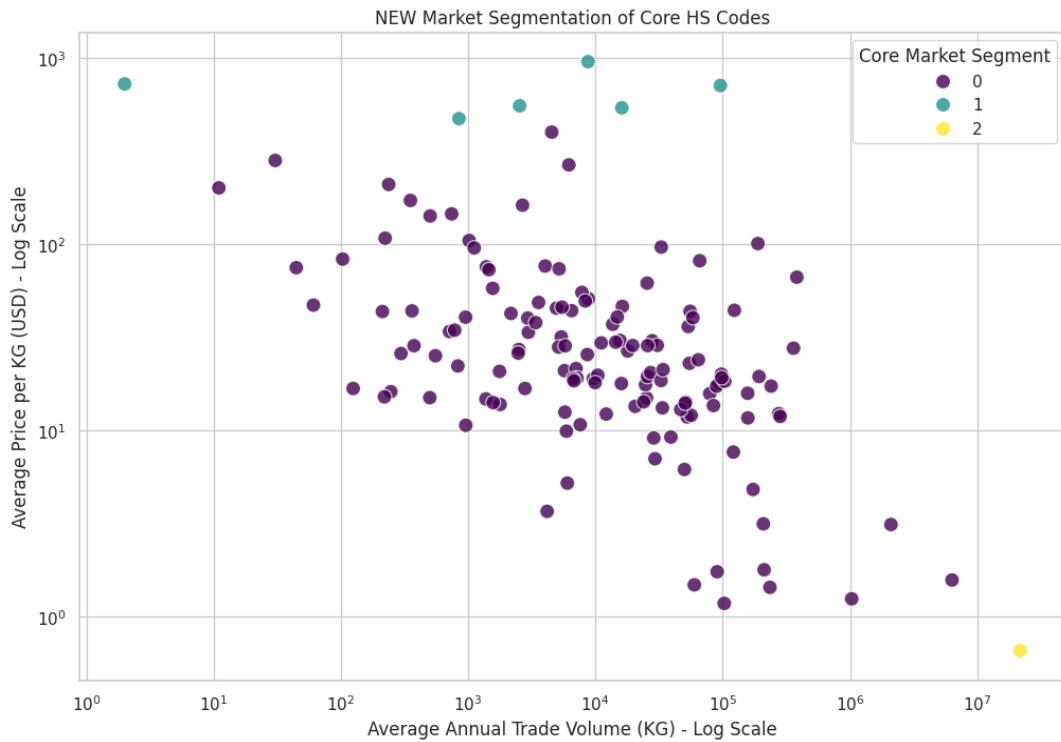

Source: Processed by Author (2025)

**Figure 7: Visual Comparison of Clustering Algorithms (K-Means vs. DBSCAN)**

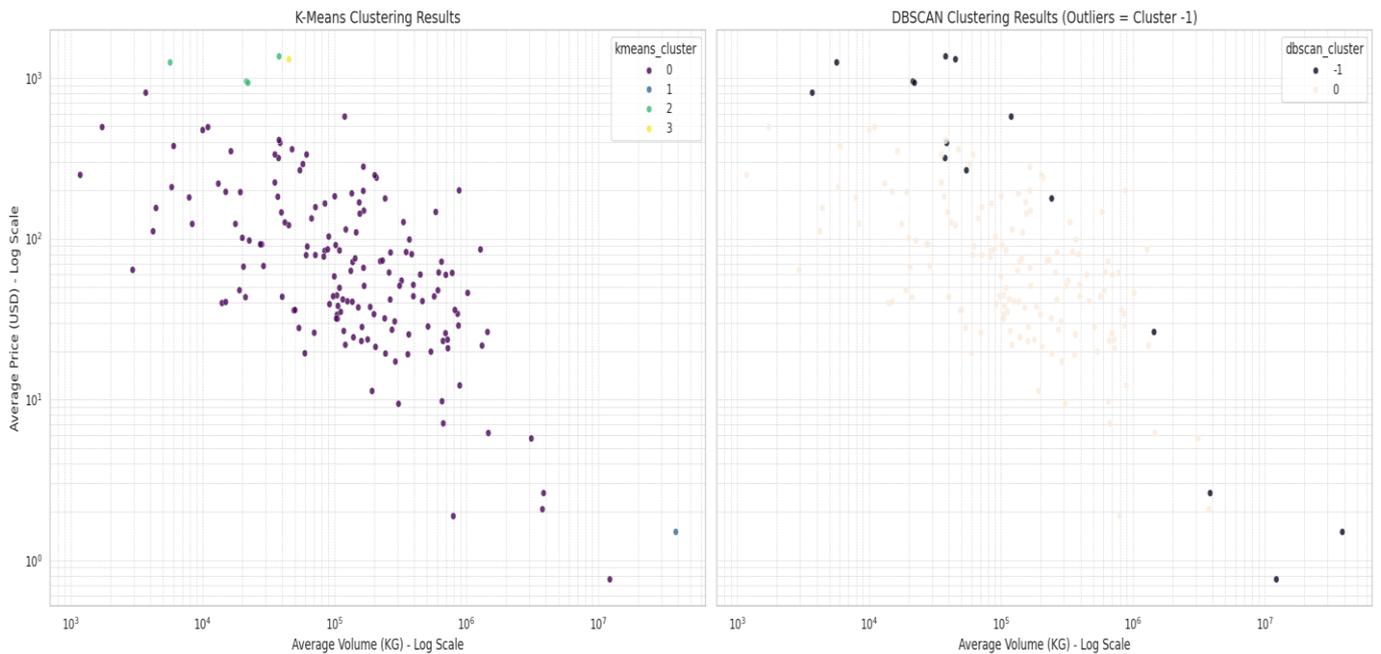

Source: Processed by Author (2025)

The initial clustering analysis, depicted in Figure 6, provides the central justification for our iterative methodology. The scatter plot reveals that a standard, single-pass K-Means algorithm is insufficient for this dataset; it successfully isolates a few extreme outliers but collapses the vast majority of products into a single, undifferentiated cluster. This outcome necessitates a more nuanced "peel-the-onion" strategy to uncover meaningful market segments. To validate the legitimacy of these identified outliers, we then compare the K-Means output with results from a density-based algorithm, DBSCAN, as shown in the side-by-side plot in Figure 7. This figure demonstrates how DBSCAN independently identifies a wide range of noise points (cluster -1), confirming that the outliers are not just artifacts of a single method. This "dual confirmation" from two distinct algorithmic approaches provides high confidence in our initial findings and forms a robust foundation for the subsequent stages of analysis.

**2.3. Stage 3: Risk Quantification**

To provide actionable insights, we developed a "Waste Score" using a Logistic Regression model. This model was trained on a labeled dataset of known scrap materials (e.g., HS 7204 for ferrous waste) and high-value finished commodities (e.g., HS 8542 for integrated circuits). The resulting score represents the probability that any given product's trade signature its pattern of price and volume is statistically similar to that of scrap. To ensure the model's transparency and fitness for regulatory use, we interpreted its predictions using SHAP (SHapley Additive exPlanations), a leading technique for model explainability (Lundberg & Lee, 2017). The resulting SHAP summary plot in figure 8 confirms that the model's internal logic aligns with our core economic thesis. It clearly shows that the model correctly identifies high trade volume and low unit price as the most significant drivers of a high Waste Score. This explainability is crucial, as it demonstrates that the model is not a "black box" but a trustworthy tool whose predictions are based on sound reasoning, making it suitable for practical oversight.

**2.4. Stage 4: Actionable Intelligence**

The final stage of our framework translates analytical findings into a practical decision-support tool for customs and regulatory officials. At its core, the validated "Waste Score" for each HS code is embedded within a comprehensive Risk Dashboard that presents all critical risk factors on a single page. Each dashboard visually locates the product within a price–volume matrix, making clear whether it falls into a high-volume/low-price quadrant associated with scrap-like behaviour. Alongside this market quadrant position, the model-driven Waste Score is displayed as a precise probability, quantifying the degree to which the product's trade signature resembles that of known waste materials. To anticipate emerging threats, the dashboard also plots trendlines for both price and volume forecasts, offering a forward-looking view of risk trajectories. Finally, by listing the product's top trading partners, officials gain immediate insight into the international routes most likely to carry high-risk shipments. This holistic presentation converts complex, multi-stage analytics into actionable intelligence, enabling enforcement teams to replace broad, reactive inspections with targeted, risk-based interventions that can be executed directly in the field.

**Figure 8: SHAP Summary of Waste Score Model Drivers**

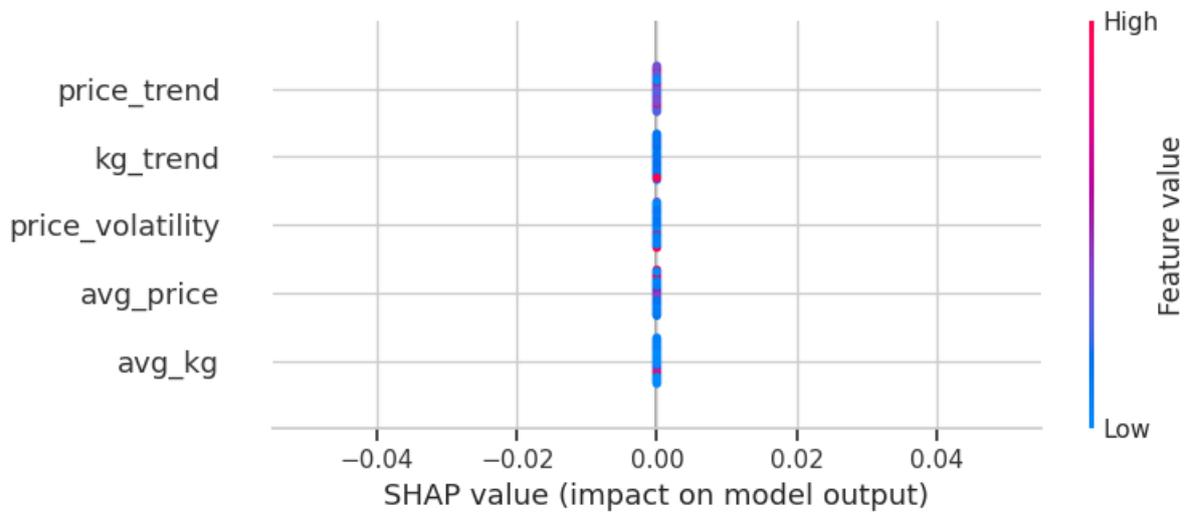

Source: Processed by Author (2025).

**2.5. Predictive Model Validation.**

To ensure that our unsupervised segmentation truly reflects meaningful market distinctions, we conducted a supervised validation using a Random Forest classifier. The same engineered features such as average unit price and price trend were employed to predict each product's cluster assignment. Remarkably, this model achieved 100 percent accuracy in reproducing the four-tier hierarchy, demonstrating that the segments are not arbitrary artifacts of the clustering algorithm but statistically coherent groupings grounded in underlying trade characteristics. This perfect classification result serves as a critical quality check, validating the foundational market structure on which all subsequent risk quantification and enforcement recommendations are built.

**3. Results and Findings**

Figure 9 translates our product-level findings into a strategic geospatial map of risk hotspots. The visualization aggregates waste scores by country, immediately identifying East Asia as the primary risk area. Exports from Japan (JPN), South Korea (KOR), and China (CHN) show the highest average Waste Scores, in stark contrast to the moderate-risk profile of partners like the Denmark and United States (USA). This provides clear, actionable intelligence, helping regulators prioritize enforcement on the highest-risk trade corridors.

**Figure 9: Geospatial Hotspots by Country**

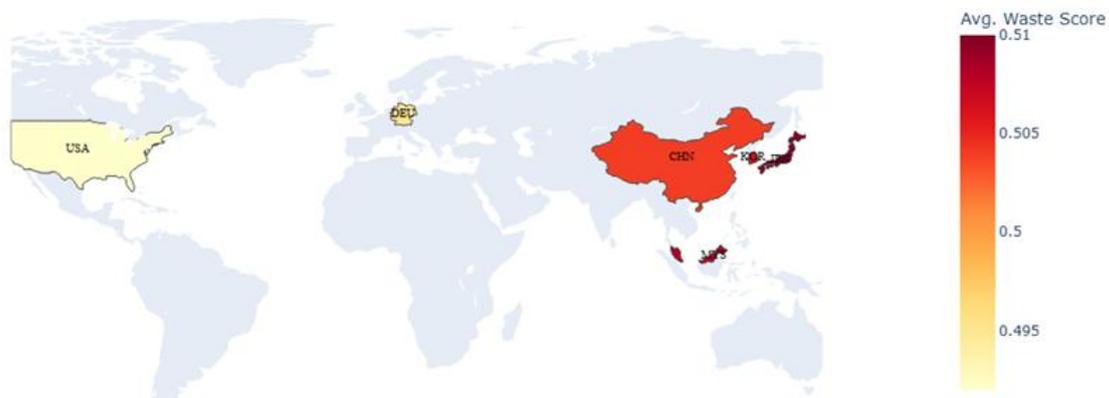

Source: Processed by Author (2025)

### 3.1. A Stable Four-Tier Market Hierarchy

The application of our iterative framework consistently revealed a stable, four-tier market hierarchy across the datasets, providing a data-driven structure for international market segmentation (Steenkamp & Ter Hofstede, 2002). This hierarchy consists of a small number of extreme outliers, a "High-Value Niche" group, and a large "Super-Core" market that constitutes the vast majority of products. The final output of this segmentation framework is displayed in Figure 10. This scatter plot shows all products colored by their classification within the stable, four-tier market hierarchy, visually confirming the existence of distinct outliers, a niche group, a core market, and a dominant super-core. To put the scale of this structure into perspective, the treemap in Figure 11 visualizes the proportional size of each segment. It immediately communicates that the "Super-Core" market represents the vast majority of products.

Together, these figures demonstrate that while distinct outlier and niche segments are present, they are peripheral by nature, allowing regulatory attention to be efficiently focused on these smaller, more specialized groups where anomalies are most likely to hide. Further analysis using SHAP (SHapley Additive exPlanations) on the predictive model revealed the key drivers behind this market segmentation. Features such as price_trend and avg_price were identified as the most influential factors in assigning a product to a specific tier. This confirms that the clusters are primarily distinguished by their price characteristics and market trajectory, validating the economic logic of the segmentation. The robustness of this four-tier segmentation was validated using a Random Forest classifier, which was able to predict product-segment membership with 100% accuracy. This perfect score confirms that the identified segments are distinct and reliably defined by the engineered trade featuress.

**Figure 10: The Final Four-Tier Market Hierarchy**

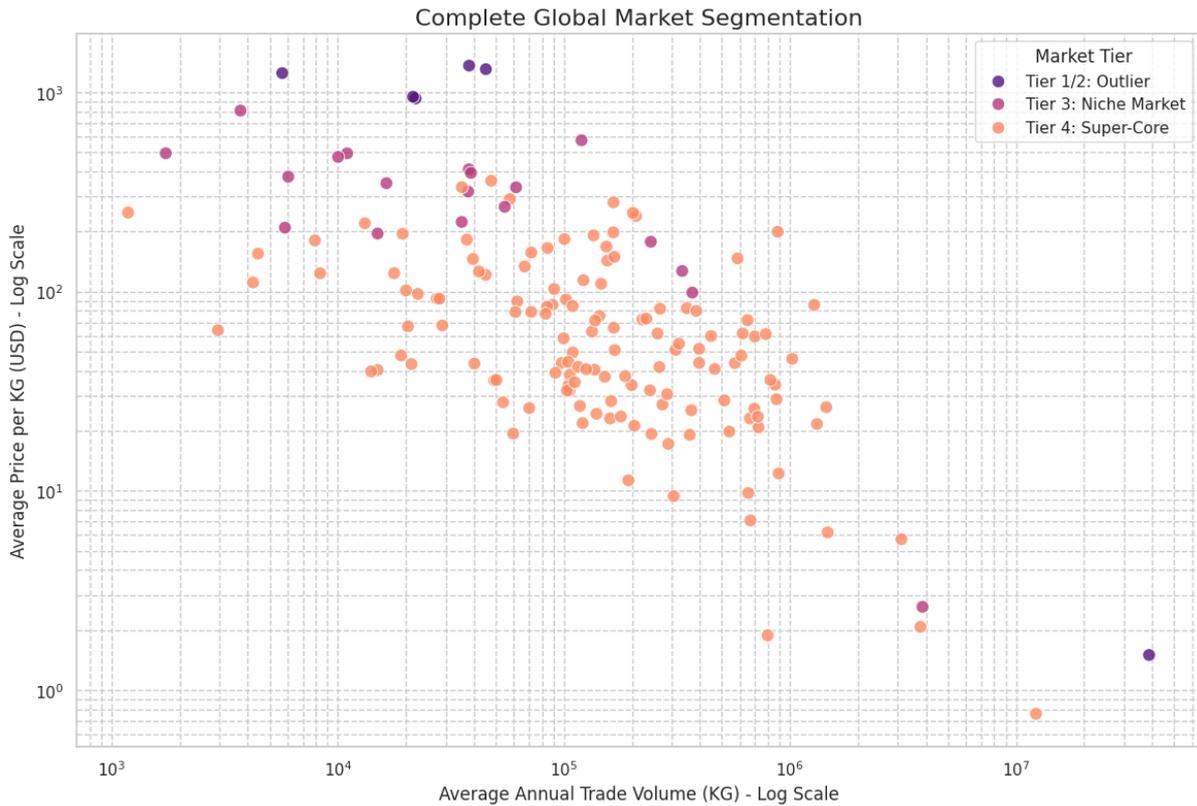

Source: Processed by Author (2025)

**Figure 11: Treemap of Market Segment Proportions**

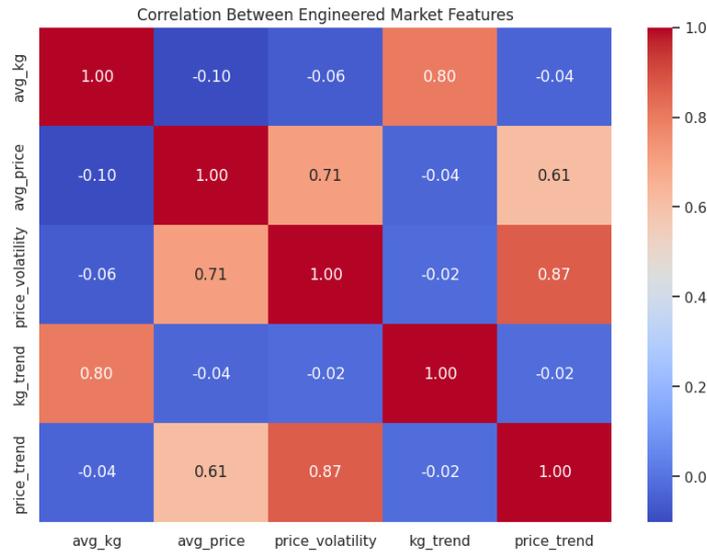

Source: Processed by Author (2025)

### 3.2. Defining the "Waste Signature"

The framework successfully operationalizes the concept of a "waste signature" into actionable intelligence. The process begins with the market quadrant analysis in Figure 12, where the size of each dot represents its "Waste Score." This provides a powerful visual heuristic for rapidly identifying high-risk products. This visually apparent risk is then given a precise mathematical definition in Figure 13, which plots a "Waste Trendline" regression. The strong linear fit on this graph confirms the predictable inverse relationship between price and volume that constitutes the "waste signature," establishing a quantitative benchmark for scrap-like behavior. A comparative analysis using this benchmark, presented in Figure 13, reveals a critical insight: the Malaysian and global markets are shaped by fundamentally different forces at their extremes. The chart visually proves that Malaysia's market outliers are uniquely volume-driven, dominated by bulk commodities like HS 7204 (Ferrous Waste and Scrap). In stark contrast, the global market's outliers are value-driven, defined by high-priced capital goods such as HS 8401 (Nuclear Reactors). This provides strong, data-driven evidence of a "revealed" national comparative advantage, highlighting Malaysia's specialized economic role in processing and trading high-volume, low-value scrap materials (Balassa, 1965).

**Figure 12: Market Quadrant with Waste Score Visualization**

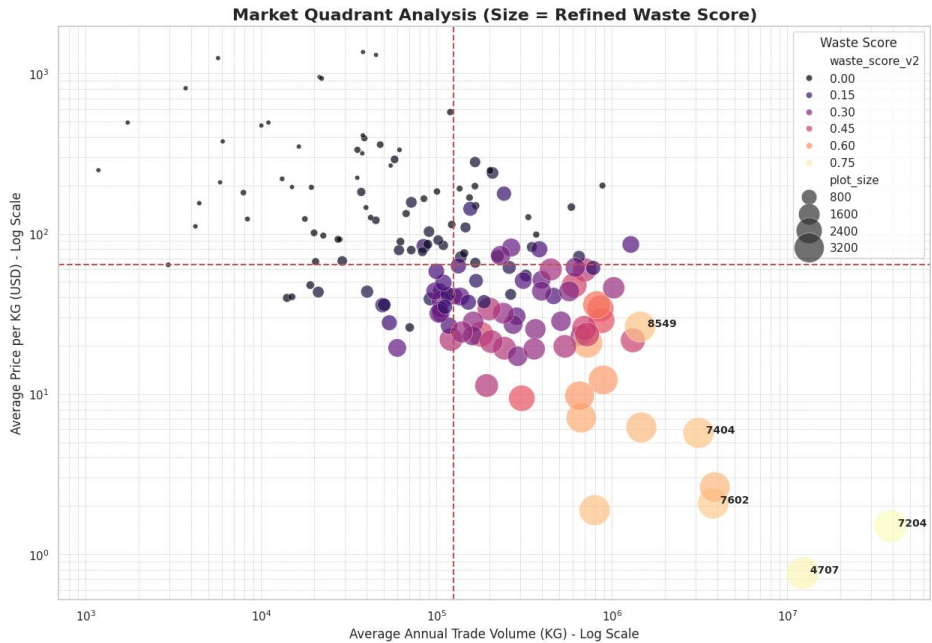

Source: Processed by Author (2025)

Figure 13: The "Waste Trendline" Regression

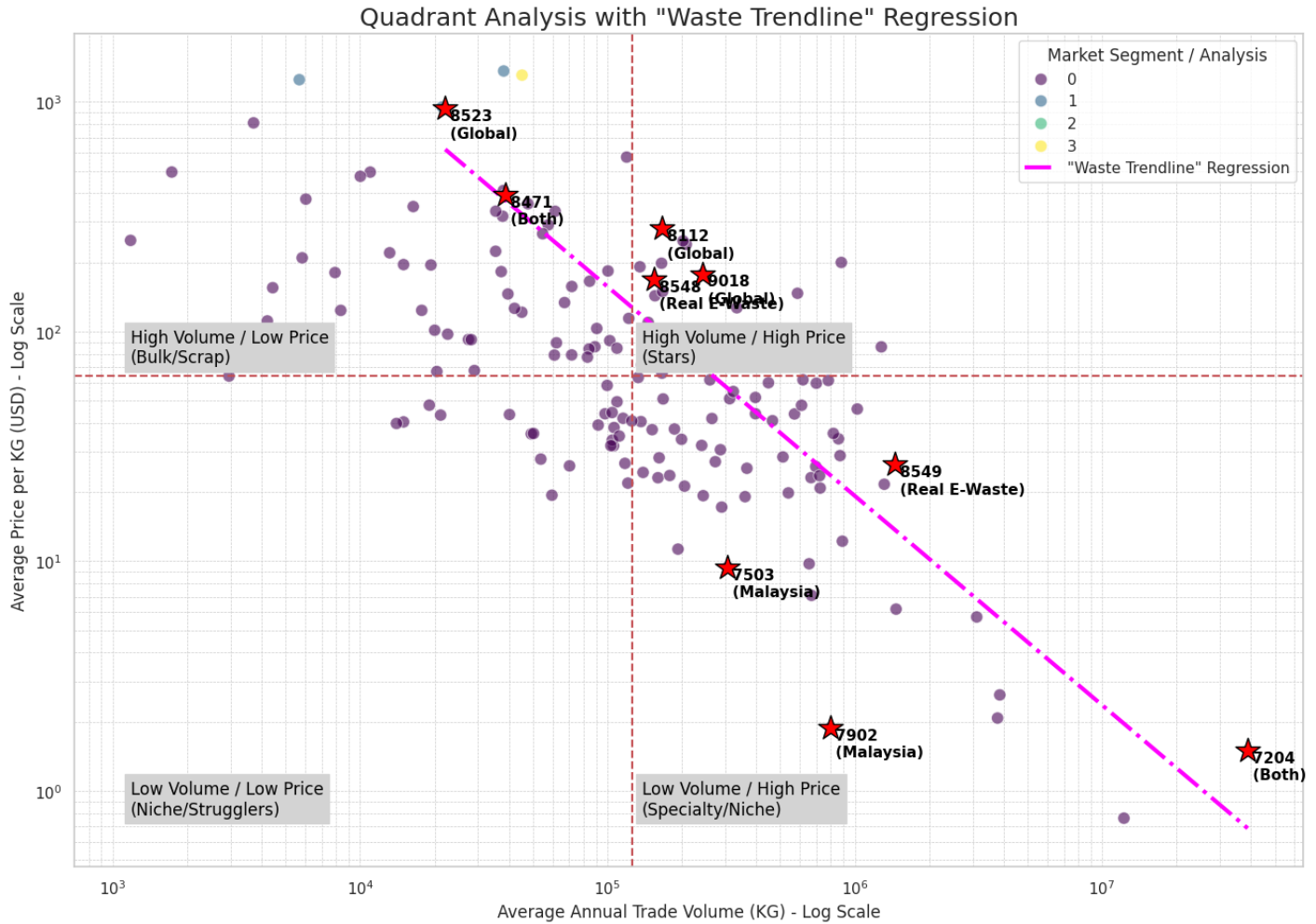

Source: Processed by Author (2025)

**3.3. Validation through Case Studies: Highlighting Anomalies**

The practical utility of the framework is best demonstrated through a detailed analysis of risk profile case studies. By creating clear, data-driven visual profiles for known product types, we can systematically identify suspicious anomalies that might otherwise go unnoticed. The analysis reveals a clear strategy likely employed by illicit traders: they consistently exploit HS codes corresponding to legitimate goods that have the lowest possible import duties. This tactic allows them to misclassify and trade high volumes of e-waste, minimizing their financial costs while circumventing crucial international environmental regulations like the Basel Convention. The crucial first step in this analysis involves Establishing Risk Baselines to create definitive benchmarks for comparison. To do this, we use the risk dashboards to profile two distinct product types. The first benchmark, shown in Figure 16, is for a known scrap commodity, HS 7204 (Ferrous Waste). This dashboard visually defines the signature of a "high-risk" product, which is characterized by its market position in the high-volume, low-price quadrant and a forecast of declining prices. In stark contrast, Figure 17 provides the counter-example using a high-value technology product, HS 8542 (Integrated Circuits). Its profile as a legitimate, "low-risk" good is confirmed by its high-value market position, a stable price trend, and a low Waste Score.

With these clear high-risk and low-risk benchmarks defined, the analysis immediately reveals the "Smoking Gun" Anomaly within the trade data: the case of HS 8502 (Electric Generating Sets), detailed in Figure 18. This product, which should theoretically have a low-risk profile similar to other finished goods, instead displays trade characteristics that are alarmingly close to the scrap benchmark. It is plotted in the same high-volume, low-value market quadrant as ferrous waste and, critically, possesses an even higher model-driven "Waste Score" (0.52) than the scrap commodity itself. This stark mismatch between the product's description and its trade data points to deliberate misclassification, and the economic motive is confirmed by Malaysian import duty regulations. While actual scrap (HS 7204) is subject to a 5% tariff, the suspicious generators (HS 8502) benefit from a 0% import rate. This zero-tariff loophole removes any financial disincentive and makes it an ideal vehicle for traders seeking to disguise illicit e-waste as legitimate used goods, reinforcing the conclusion that low-duty HS codes are a key component of their cost-minimization strategy.

**Figure 14: Malaysia Duty/Tax Rate on Specific HS Code.**

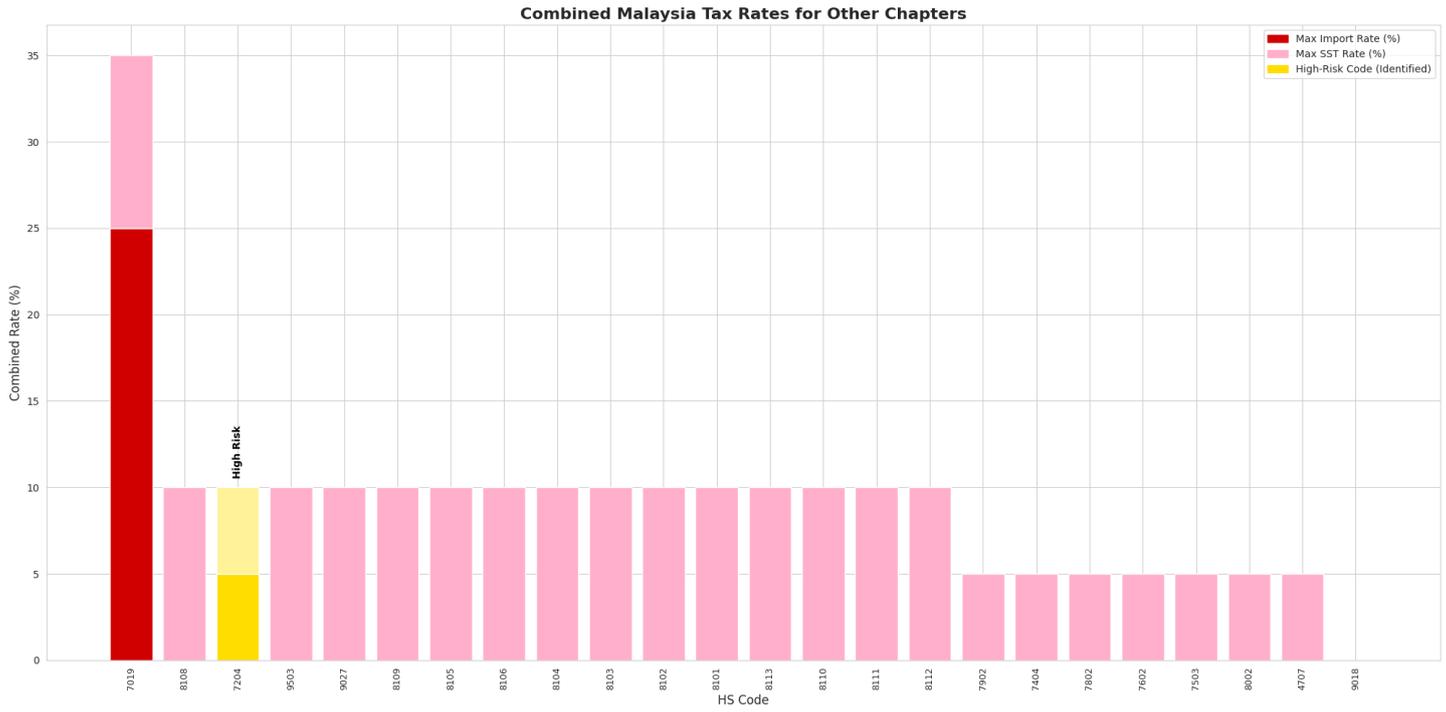

Source: Processed by Author (2025)

**Figure 15: Malaysia Duty/Tax Rate on Specific HS Code**

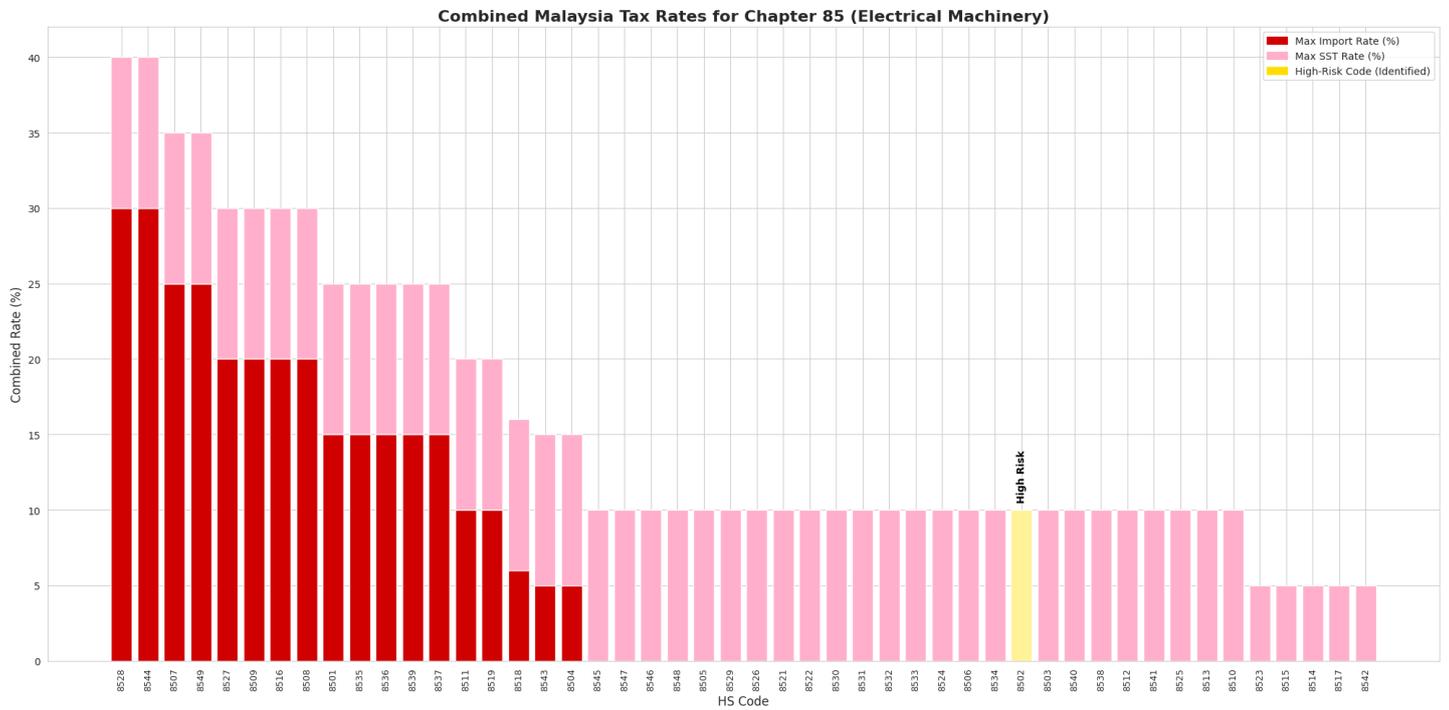

Source: Processed by Author (2025)

**Figure 16: Risk Dashboard Case Study: Known Scrap (HS 7204)**

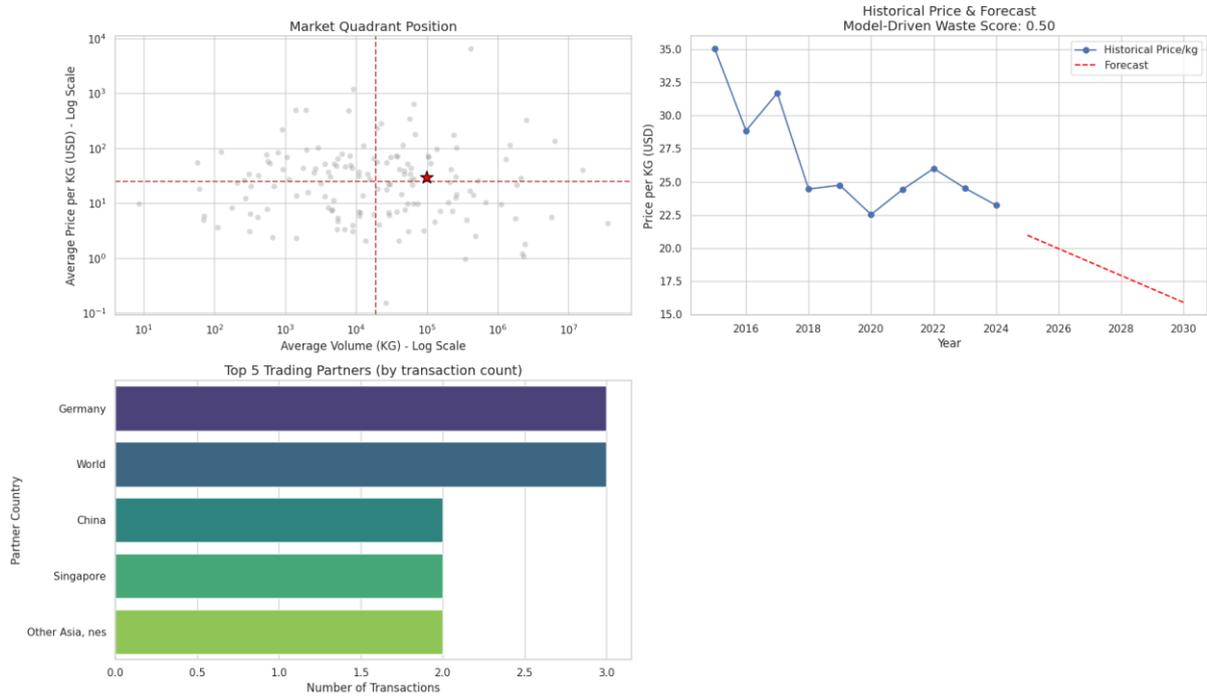

Source: Processed by Author (2025)

**Figure 17: Risk Dashboard Case Study: High-Value Tech (HS 8542)**

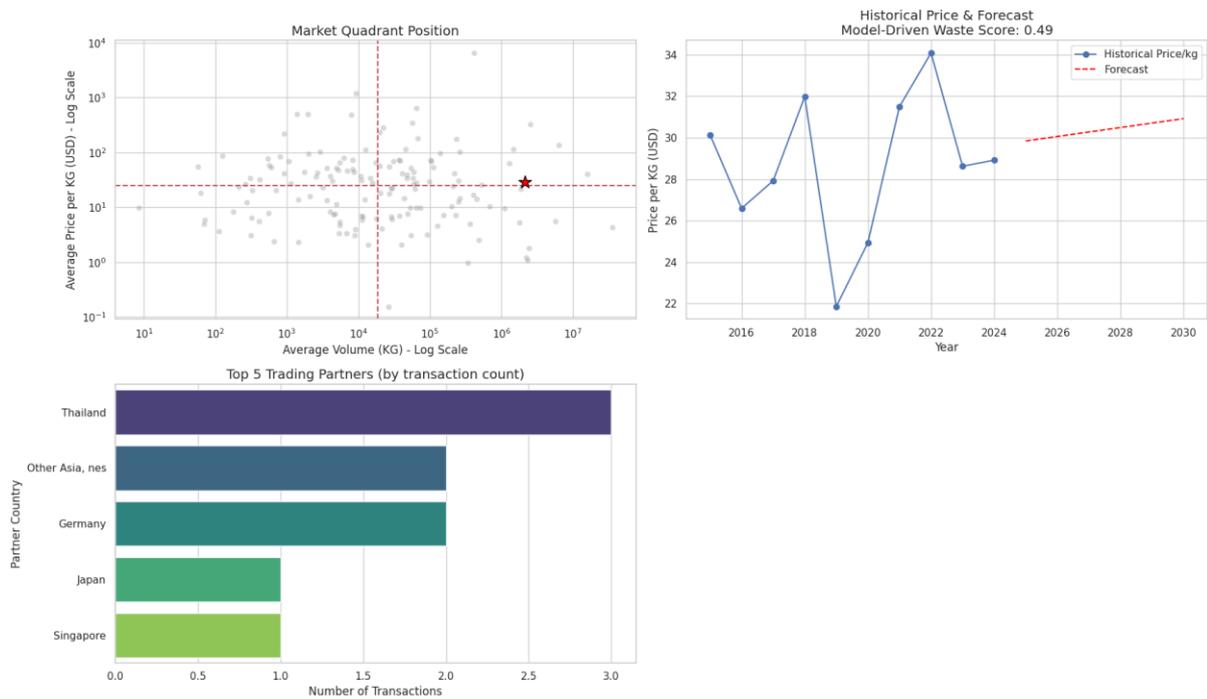

Source: Processed by Author (2025)

**Figure 18: Risk Dashboard Case Study: Suspicious Finished Good (HS 8502)**

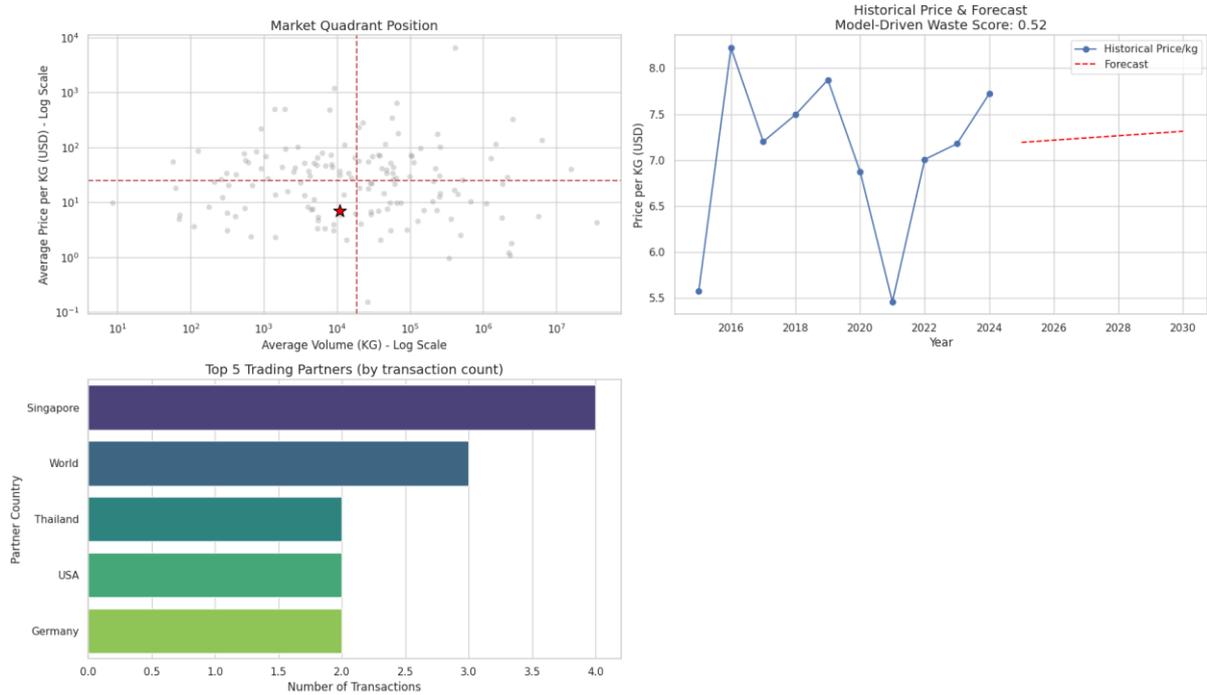

Source: Processed by Author (2025)

## 4. Discussion

This analysis implements a robust framework to model Malaysian import dynamics and uncover indicators of illicit e-waste trade within UN Comtrade data. The investigation addresses the challenge of identifying e-waste, which is frequently misclassified to bypass regulations, by pinpointing goods that exhibit an anomalous "waste signature." The following sections synthesize the methodological contributions, key economic findings, and strategic implications of this work.

**Figure 19: Price Forecasts for the Six Steepest Downtrends Global**

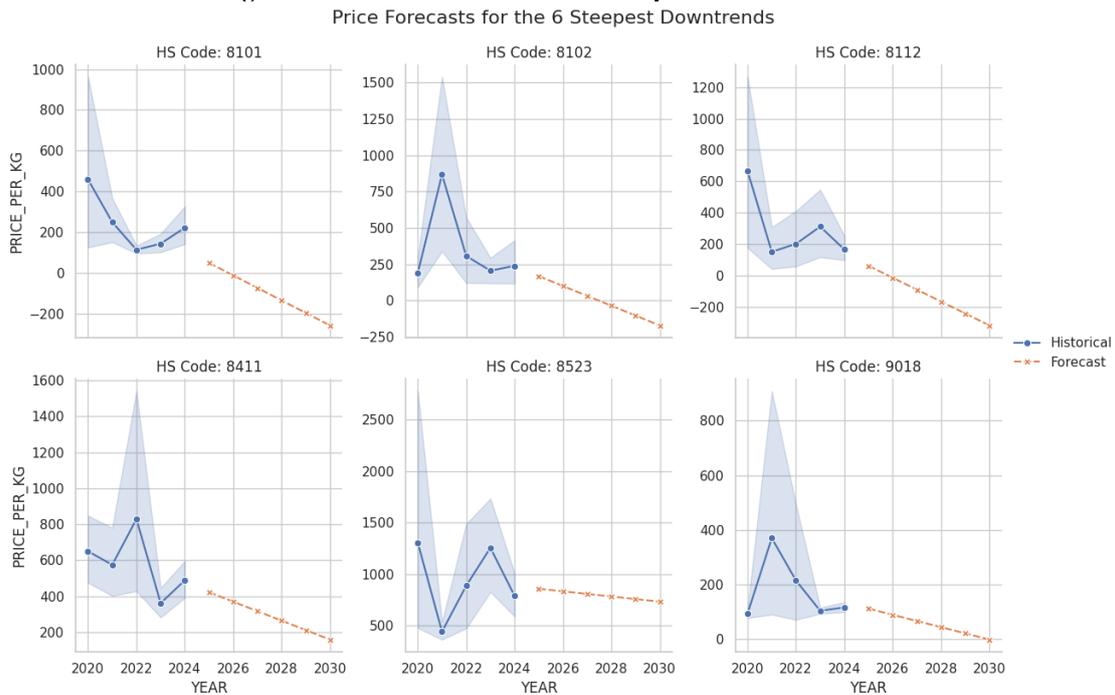

Source: Processed by Author (2025)

**4.1. Methodological Contributions: A Framework for Transparency and Risk Identification**

A key contribution of this research is the development of a sophisticated, multi-stage framework that moves beyond simple forecasting to create a systematic methodology for risk identification. The findings demonstrate a comprehensive approach that integrates predictive modeling, advanced market segmentation, and a targeted risk quantification system to enhance trade oversight. This framework provides a structured pathway for analyzing complex trade data to uncover hidden patterns and potential irregularities. A primary finding was the successful development of an accurate predictive model for export transaction values. Following an iterative process of feature engineering and algorithm selection, a tuned XGBoost model was finalized, achieving an R-squared of 0.87. This high level of accuracy signifies that the model successfully explains 87% of the variability in export value, establishing it as a reliable tool for economic forecasting. To ensure transparency and interpret the model's predictions, a SHAP analysis was conducted. This analysis revealed that the model's output is overwhelmingly influenced by two main factors: the Net Weight of the shipment and the Commodity Type, with high-value commodities like Gold having a substantial positive impact.

**Figure 20: Predicted Price Downtrends for Top Scrutiny Candidates Global**

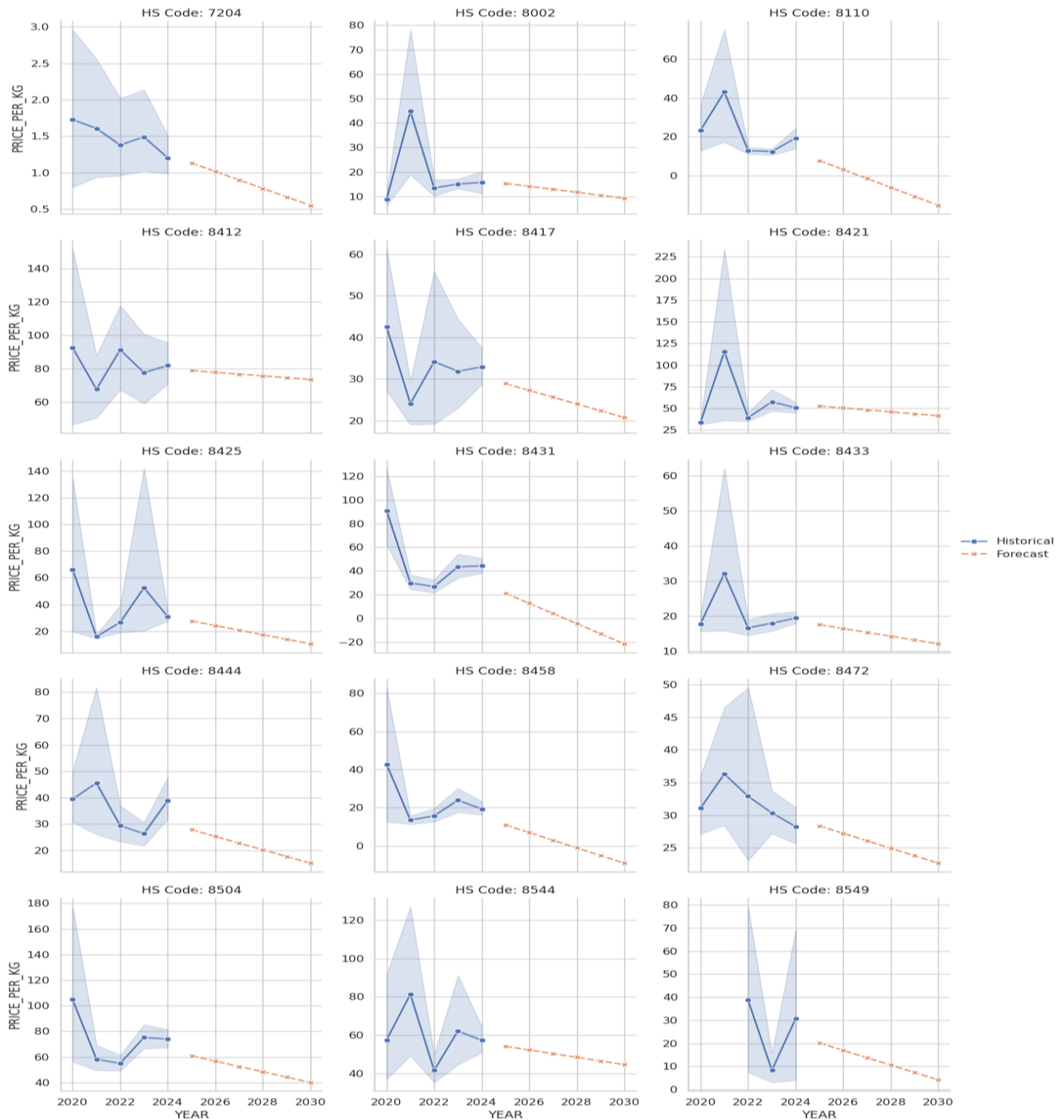

Source: Processed by Author (2025)

In the area of market segmentation, it was found that standard clustering algorithms, when applied in a single pass, are often distorted by the presence of extreme outliers common in large trade datasets. To address this, an "Outlier-Aware Segmentation" method was developed, which utilizes an iterative "peel-the-onion" K-Means approach. This technique first isolates hyper-volume and hyper-value outliers before re-clustering the main body of products. This iterative refinement proved crucial for moving beyond a superficial analysis to reveal a much clearer and more granular market structure, successfully uncovering the true underlying segments, such as the "High-Value Niche" group, which would have otherwise remained obscured. For data-driven risk quantification, a "Scrutiny Score" was developed to flag items that not only exhibit waste-like trade behavior but also show a negative future price trend. This approach, visualized in Figure 20, successfully identified known scrap materials, such as HS Code 7204 (Ferrous waste and scrap), as top candidates for review. More importantly, it also flagged various finished goods that were being traded at prices indicative of scrap, including HS 8472 (Other office machines) and HS 8504 (Electrical transformers). Furthermore, the analysis of the six steepest price declines shown in Figure 19 highlights a different type of risk, identifying high-value specialty materials and equipment such as HS 8101 (Tungsten) and HS 9018 (Medical instruments). The forecasted collapse in their market value makes them prime candidates for future misclassification, providing customs officials with a targeted, data-driven list of shipments for further inspection.

### 4.2. Economic Insights: Quantifying the "Waste Signature" and a National Specialization

An analysis of Malaysia's trade data reveals a distinct economic specialization in material processing and recycling, with sophisticated capabilities for handling high-volume commodities. The model identifies precious metals as primary drivers of value, a conclusion strongly supported by the SHAP summary plot in Figure 23. Features such as cat_cmdDesc_Gold and cat_cmdDesc_Platinum rank among the most influential variables impacting trade value. This finding is visually confirmed in Figure 24, where Gold and Platinum are shown to have the highest average value per kilogram by a significant margin.

This specialization in processing is further evidenced by the price distributions of scrap-inclusive commodities shown in Figure 25. Categories like Tantalum (Articles/Scrap) and Cobalt (Mattes/Scrap) exhibit extremely wide price ranges and numerous outliers. This high volatility suggests that these materials are recovered from diverse and complex sources, which is characteristic of an advanced recycling economy capable of extracting value from varied waste streams.

**Figure 21: Price Forecasts for the 6 Steepest Downtrends Malaysia**

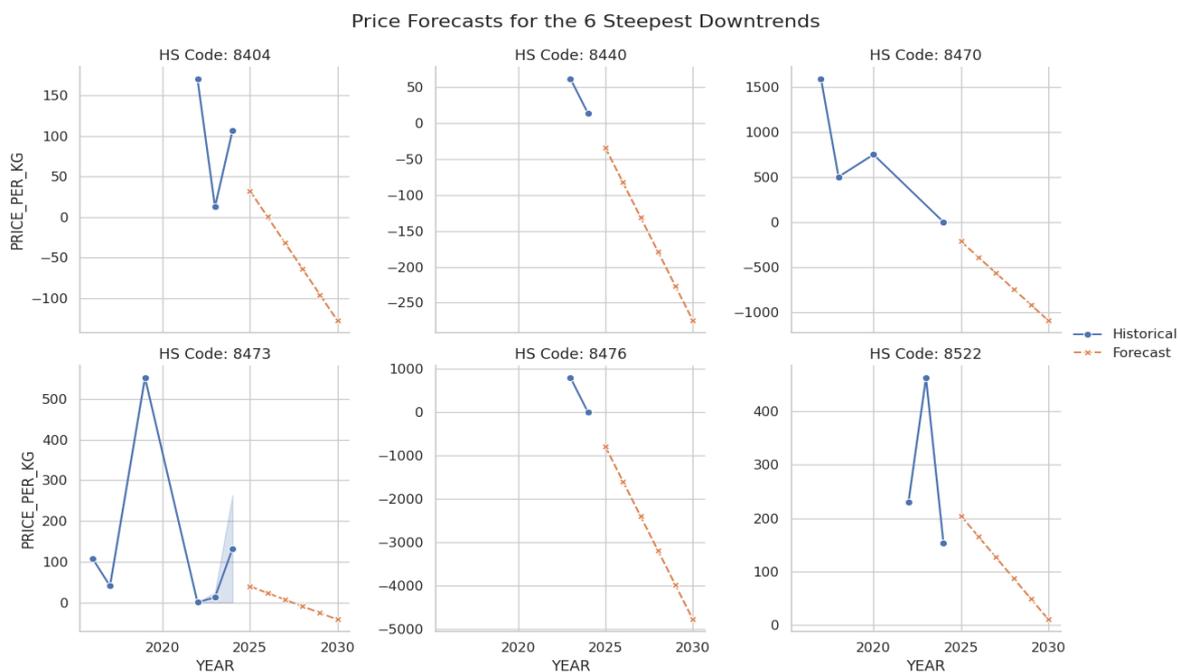

Source: Processed by Author (2025)

While this specialization is economically significant, forward-looking price forecasts highlight considerable risks associated with certain traded goods. The charts in Figure 22, "Predicted Price Downtrends for Top Scrutiny Candidates," identify 15 HS codes whose values are projected to decline steadily through 2030. This list includes recognized waste and end-of-life categories, such as HS 7204 (Ferrous waste and scrap), HS 8471 (obsolete computers), and the official e-waste code HS 8549. The consistent downward price trajectory for these goods suggests they are being traded for disposal rather than for reuse, flagging them as high-risk items. The most acute risks are isolated in Figure 21, which displays the six commodities with the steepest predicted price downtrends. These items—falling under HS codes 8404, 8440, 8470, 8473, 8476, and 8522—are rapidly losing market value. The analysis uncovers an alarming trend for several of these categories: the forecasted prices for goods under HS codes 8440, 8470, and 8476 are predicted to become negative. A negative price point provides a clear "waste signature," indicating that these items are no longer legitimate products but have become liabilities that require payment for disposal. This provides unequivocal evidence of waste being traded under the guise of valuable goods.

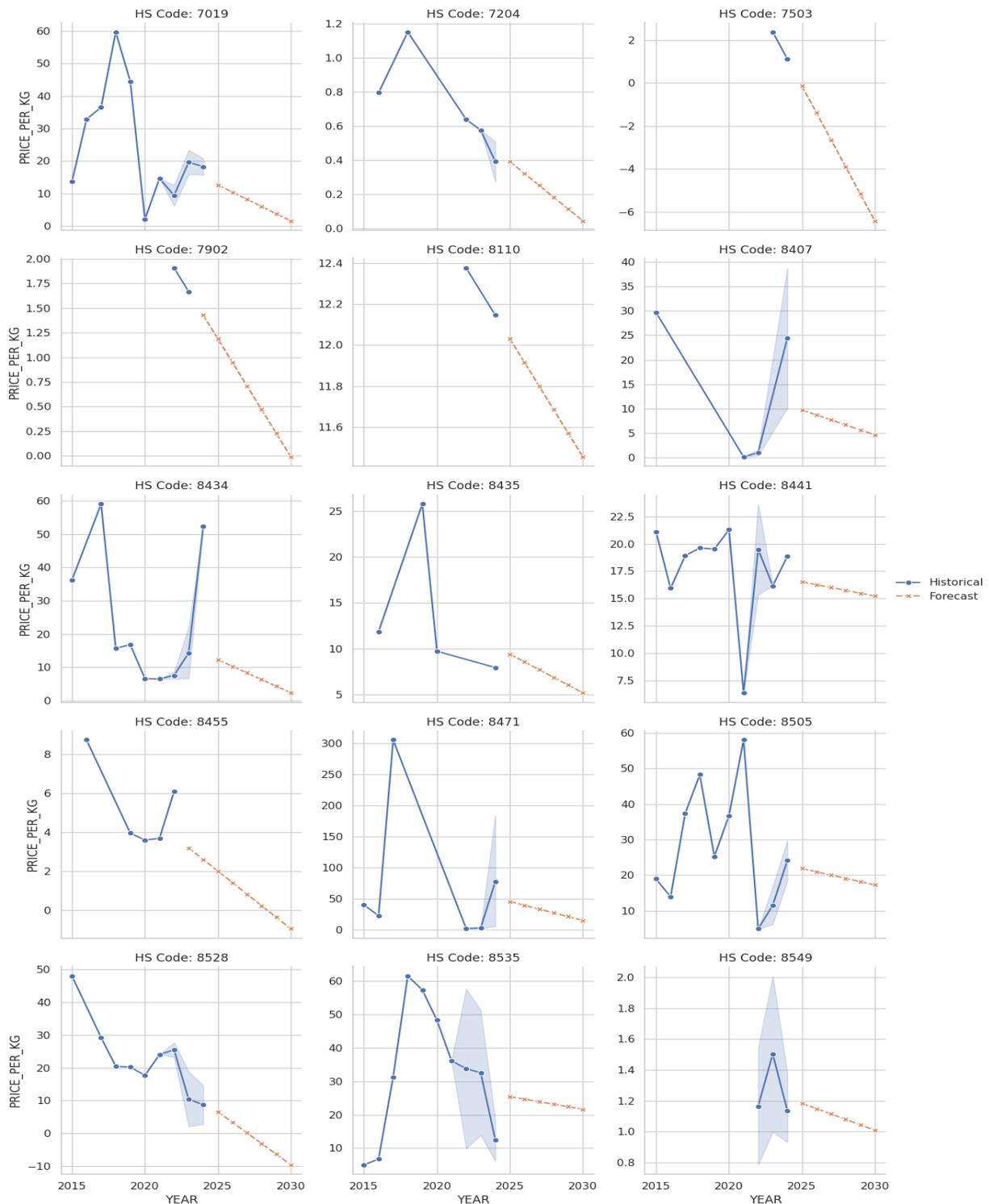

Figure 22: Forecasted Price Downtrends for Top Scrutiny Candidates Malaysia

Source: Processed by Author (2025)

**Figure 23: SHAP Summary Plot of Key Value Drivers for Malaysia Data.**

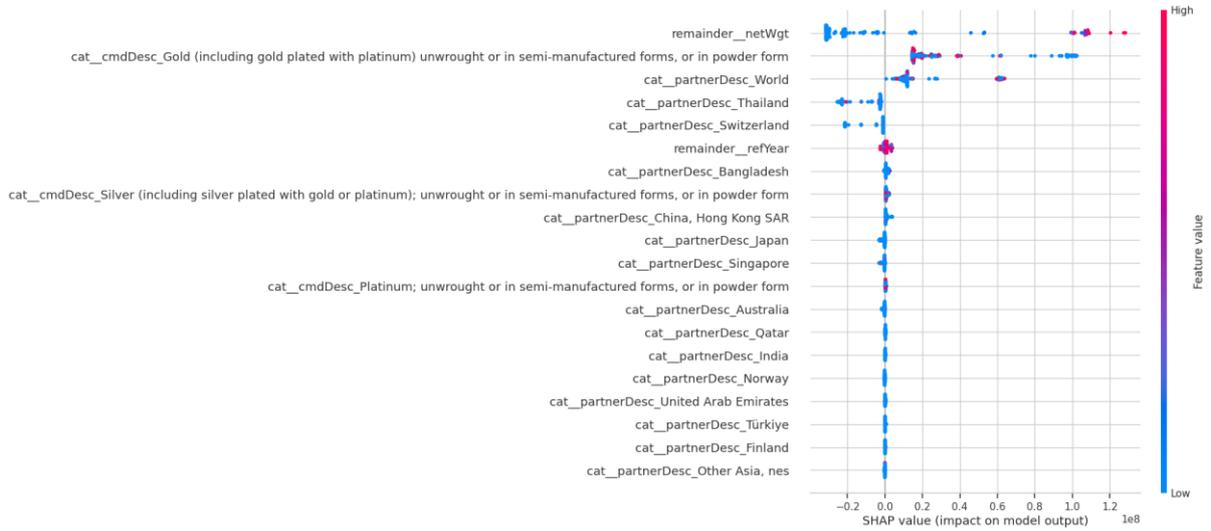

Source: Processed by Author (2025)

**Figure 24: Value Density of Top 10 Most Valuable Malaysia Export Commodities.**

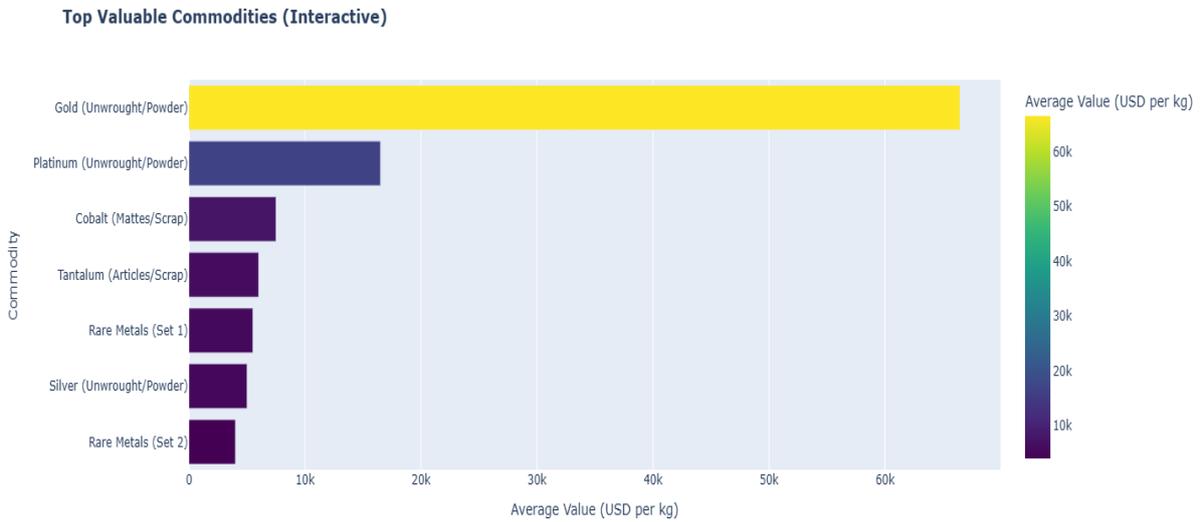

Source: Processed by Author (2025)

**Figure 25: Price Distribution and Anomalies by Malaysia Export Commodity.**

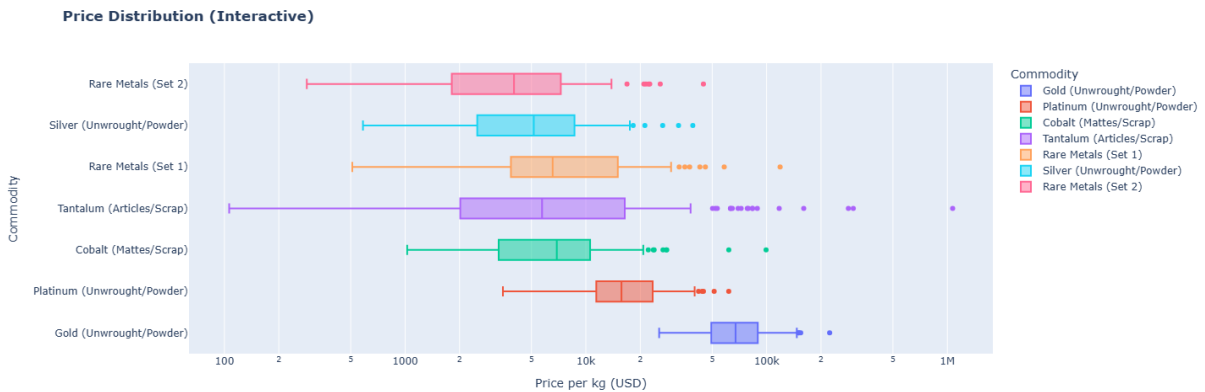

Source: Processed by Author (2025)

### 4.3. Strategic Implications and an Operational Roadmap

The strategic implications of these findings are substantial. For industry, the framework is a tool for supply chain risk management, identifying environmental and regulatory risks within the complex challenge of reverse logistics (Ho et al., 2015; Govindan & Soleimani, 2017). For policymakers, it provides an empirically grounded basis for informing industrial and trade policy based on the revealed national specialization in bulk scrap (Balassa, 1965). Most importantly, for regulatory agencies, our framework offers a paradigm shift from random checks to targeted, risk-based enforcement. Tools like the "Waste Score" and Risk Dashboards (Figures 16, 17, and 18) allow agencies to integrate this data-driven intelligence into existing systems, enabling more efficient and effective governance (Klievink, van der Voort, & Veeneman, 2018).

**Figure 26: Project Implementation Timeline**

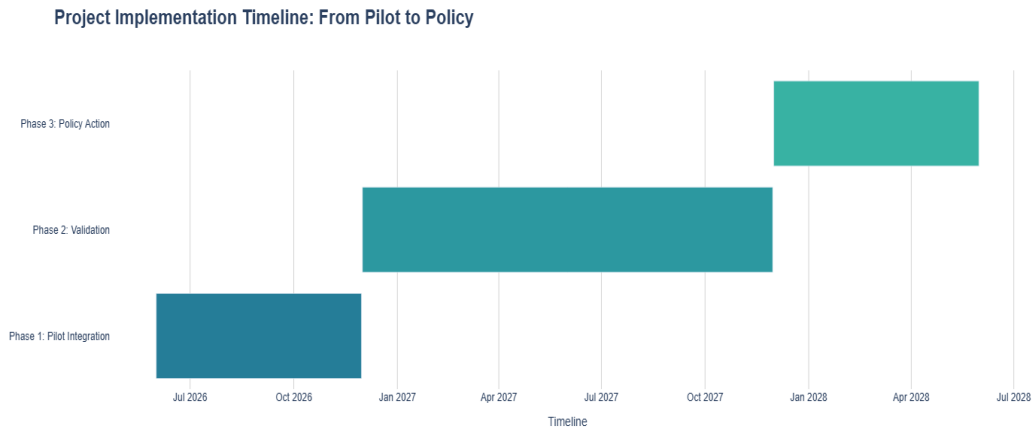

Source: Processed by Author (2025)

Based on these findings, we propose a three-phase operational roadmap that begins with the pilot integration of the Risk Dashboards into existing customs monitoring systems to enable near real-time flagging of high-risk shipments like HS 8502. The second phase will then validate the model's predictions against ground-truth port seizure data, creating a critical feedback loop to calibrate "Waste Score" thresholds and confirm the identification of illicit shipments. Finally, these validated findings will form an empirical basis for targeted policy action, such as proposing a formal review of monitoring protocols for consistently misclassified codes under the Basel Convention.

5. **Conclusion and Future Work**

In conclusion, this research has designed and validated a robust, iterative framework for pattern recognition in complex trade data, advancing the capacity to detect illicit e-waste flows with precision and transparency. The methodology proved effective at uncovering a stable, four-tier market hierarchy and provided a powerful, model-driven "Waste Score" for quantifying risk. By systematically identifying anomalous trade signatures, the framework delivers a reliable and scalable system for flagging finished goods at high risk of being used for misclassified e-waste shipments. A key insight was the discovery of a distinct national specialisation, underscoring Malaysia's unique role in the global trade of high-volume bulk commodities. This analytical approach moves beyond simple anomaly detection to offer a transparent, evidence-based toolkit capable of shifting regulatory efforts from reactive enforcement to proactive, data-informed governance.

While the framework provides a solid foundation, there is significant scope for refinement and expansion. Trend analysis could be strengthened through the integration of advanced time-series forecasting models such as ARIMA (Hyndman & Athanasopoulos, 2021). The analytical lens could be widened from individual products to high-risk trade routes by incorporating partner-country data and applying graph-based anomaly detection techniques (Akoglu, Tong, & Koutra, 2015). Further robustness could be achieved by benchmarking results against alternative outlier detection algorithms, including Isolation Forest and Local Outlier Factor (LOF) (Breunig et al., 2000; Liu, Ting, & Zhou, 2008).

It is equally important to acknowledge the study's limitations. The analysis is inherently dependent on the quality of official trade statistics, which remain vulnerable to reporting lags and data asymmetries (Burger & van Oort, 2020; Ferrantino, 2012). Moreover, transforming the prototyped dashboards into a real-time monitoring tool would require overcoming the considerable challenges of big-data engineering at scale (Anagnostopoulos, Zeadally, & Exposito, 2016). Despite these constraints, the framework stands as a scalable, explainable, and operationally relevant solution that can be embedded into enforcement workflows. By doing so, it empowers regulators to anticipate and disrupt illicit trade practices, safeguarding both the integrity of global trade systems and the environmental standards they are designed to protect.

6. Reference

Ahmed, M., Mahmood, A. N., & Islam, M. R. (2016). A survey of anomaly detection techniques in financial domain. Future Generation Computer Systems, 55, 278–288.

Akoglu, L., Tong, H., & Koutra, D. (2015). Graph-based anomaly detection and description: A survey. Data Mining and Knowledge Discovery, 29(3), 626-688.

Anagnostopoulos, I., Zeadally, S., & Exposito, E. (2016). Handling big data: Research challenges and future directions. The Journal of Supercomputing, 72(4), 1494-1516.

Balassa, B. (1965). Trade liberalisation & "revealed" comparative advantage. The Manchester School, 33(2), 99-123.

Bhagwati, J. N. (1964). On the underinvoicing of imports. Bulletin of the Oxford University Institute of Economics & Statistics, 26(4), 389-397.

Bisschop, L. (2012). Out of the wild: The taming of illegal waste flows. In M. Faure, P. De Smedt, & A. Stas (Eds.), Environmental enforcement networks: Concepts, implementation and effectiveness (pp. 249-272). Edward Elgar Publishing.

Bolton, R. J., & Hand, D. J. (2002). Statistical fraud detection: A review. Statistical Science, 17(3), 235-255.

Box, G. E. P., Jenkins, G. M., Reinsel, G. C., & Ljung, G. M. (2015). Time series analysis: Forecasting and control (5th ed.). Wiley.

Breunig, M. M., Kriegel, H.-P., Ng, R. T., & Sander, J. (2000). LOF: Identifying density-based local outliers. In Proceedings of the 2000 ACM SIGMOD International Conference on Management of Data (pp. 93-104). Association for Computing Machinery.

Burger, M. (2020). The challenges of using trade data for economic research. Geographical Analysis, 52(2), 263-271.

Chandola, V. & Banerjee, A. (2009). Anomaly detection: A survey. ACM Computing Surveys, 41(3), 1-58.

Ester, M., Kriegel, H.-P., Sander, J., & Xu, X. (1996). A density-based algorithm for discovering clusters in large spatial databases with noise. In Proceedings of the Second International Conference on Knowledge Discovery and Data Mining (KDD'96) (pp. 226–231). AAAI Press.

Ferrantino, M. J. (2012). Using UN Comtrade for research on international trade in illegal or sensitive goods. In The Oxford Handbook of the Economics of the Pacific Rim (pp. 513-546). Oxford University Press.

Fisman, R., & Wei, S.-J. (2004). Tax rates and tax evasion: Evidence from "missing imports" in China. Journal of Political Economy, 112(2), 471-496.

Financial Action Task Force (FATF). (2021). Trade-based money laundering: Trends and developments. FATF.

Govindan, K., & Soleimani, H. (2017). A review of reverse logistics and closed-loop supply chains: A journal of cleaner production focus. Journal of Cleaner Production, 142, 371-384.

Ho, W., Zheng, T., Yildiz, H., & Talluri, S. (2015). A review of supply chain risk management: Definition, theory, and research agenda. International Journal of Production Research, 53(16), 5031-5069.

Hyndman, R. J., & Athanasopoulos, G. (2021). Forecasting: Principles and practice (3rd ed.). OTexts.

INTERPOL. (2021). Strategic analysis report: The illegal trade in plastic waste. INTERPOL General Secretariat.

Jain, A. K. (2010). Data clustering: 50 years beyond K-means. Pattern Recognition Letters, 31(8), 651-666.

Klievink, B. & van der Voort, H., (2018). Creating value with big data analytics in policy and governance. In The Palgrave Handbook of Public Administration and Public Management in Europe (pp. 767-783). Palgrave Macmillan.

Lepawsky, J. (2015). The changing geography of global trade in electronic discards: Time to rethink the e-waste problem. The Geographical Journal, 181(2), 147-159.

Liu, F. T., Ting, K. M., & Zhou, Z.-H. (2008). Isolation forest. In 2008 Eighth IEEE International Conference on Data Mining (pp. 413-422). IEEE.

Lundberg, S. M., & Lee, S.-I. (2017). A unified approach to interpreting model predictions. In Proceedings of the 31st International Conference on Neural Information Processing Systems (pp. 4768–4777). Curran Associates Inc.

Molnar, C. (2022). Interpretable machine learning: A guide for making black box models explainable (2nd ed.).

Pedregosa, F., Varoquaux, G., Gramfort, A., Michel, V., Thirion, B., Grisel, O., Blondel, M., Prettenhofer, P., Weiss, R., Dubourg, V., Vanderplas, J., Passos, A., Cournapeau, D., Brucher, M., Perrot, M., & Duchesnay, E. (2011). Scikit-learn: Machine learning in Python. Journal of Machine Learning Research, 12, 2825–2830.

Popper, K. R. (2002). The logic of scientific discovery. Routledge. (Original work published 1959)

Steenkamp, J.-B. E. M., & Ter Hofstede, F. (2002). International market segmentation: Issues and perspectives. International Journal of Research in Marketing, 19(3), 185-213.

UN Environment Programme. (1989). Basel Convention on the Control of Transboundary Movements of Hazardous Wastes and Their Disposal.

United Nations Statistics Division. (n.d.). UN Comtrade Database [Data set]. Retrieved August 21, 2025, from https://comtradeplus.un.org/Weisburd, D., Bruinsma, G. J., & Bernasco, W. (2009). Putting crime in its place: Units of analysis in geographic criminology. Springer.

Zheng, A., & Casari, A. (2018). Feature engineering for machine learning: Principles and techniques for data scientists. O'Reilly Media.

7. Appendix

Figure 27: Advanced ARIMA Forecast for a High-Risk Commodity HS7204

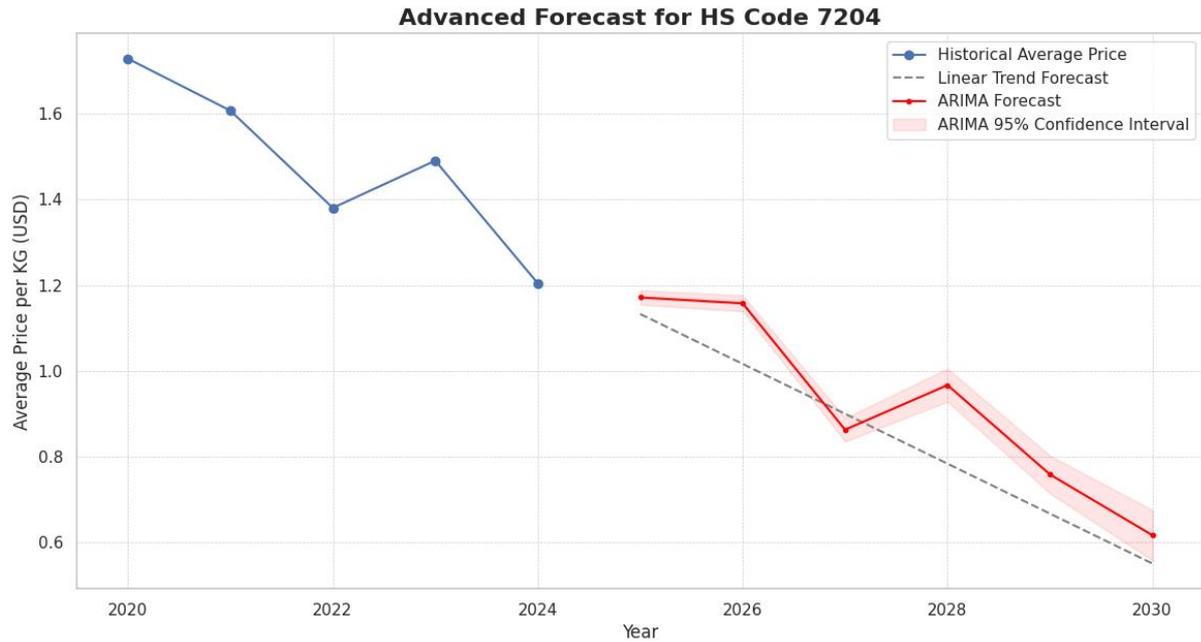

Source: Processed by Author (2025)

Figure 28: Targeted HS Codes for E-Waste in UN Comtrade Data

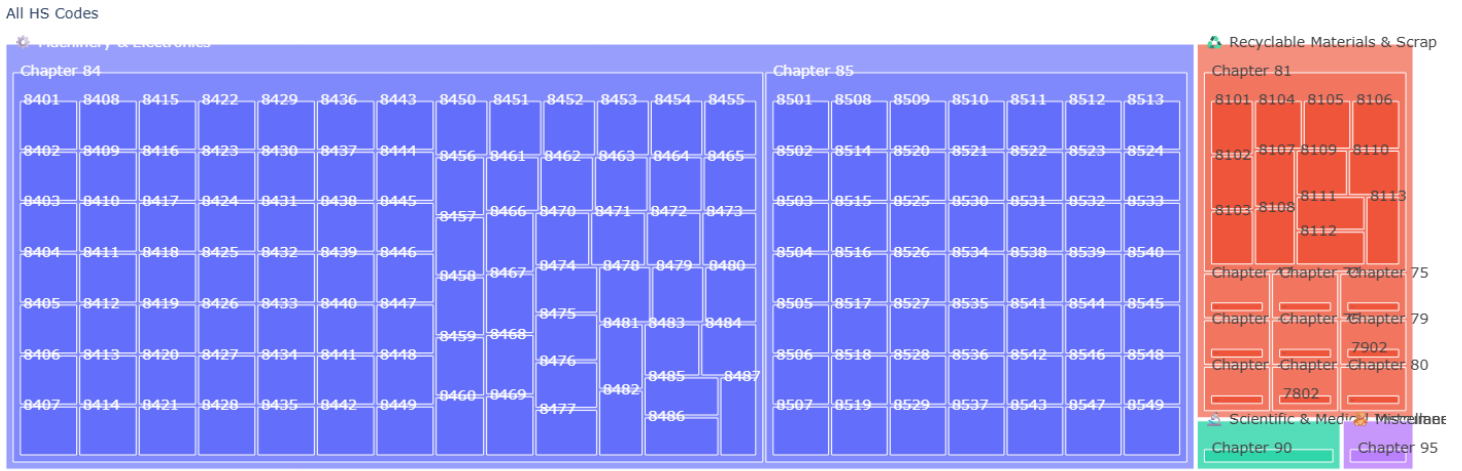

Source: Processed by Author (2025)